%% file: main.tex
\documentclass[10pt, conference]{IEEEtran}
\usepackage{cite}
\usepackage{graphicx}
\usepackage{textcomp}
\usepackage[table]{xcolor}
\usepackage[hyphens]{url}
\usepackage{amssymb, amsmath}
\usepackage[absolute,showboxes]{textpos}

\usepackage{xfrac}
\usepackage{algorithm}
\usepackage{algpseudocode}
\usepackage{xspace}
\usepackage{multirow}
\usepackage[caption=false]{subfig}
\usepackage{lipsum}
\usepackage{anyfontsize}
\usepackage{kotex}
\usepackage{tabularx}
\usepackage{booktabs}
\usepackage{enumitem}
\usepackage{comment}
\usepackage[all]{nowidow}
\usepackage[]{hyperref}
\usepackage{balance}

\newcommand{\NAME}{Theodosian\xspace}

\newcommand{\poly}[1]{\mathfrak{#1}}
\newcommand{\textpoly}[1]{$\poly{#1}$\xspace}

\TPMargin{5pt}

\newcommand{\copyrightstatement}{
    \begin{textblock}{0.84}(0.08,0.93)    
         \noindent
         \footnotesize
         \textcopyright\ 2026 IEEE. Personal use of this material is permitted. Permission from IEEE must be obtained for all other uses, in any current or future media, including reprinting/republishing this material for advertising or promotional purposes, creating new collective works, for resale or redistribution to servers or lists, or reuse of any copyrighted component of this work in other works.
    \end{textblock}
}

\begin{document}
\copyrightstatement


\title{Theodosian: A Deep Dive into Memory-Hierarchy-Centric FHE Acceleration}

\author{
\IEEEauthorblockN{
Wonseok Choi,
Hyunah Yu,
Jongmin Kim,
Hyesung Ji,
Jaiyoung Park,
and {Jung Ho} Ahn
}

\IEEEauthorblockA{Seoul National University, Seoul, Republic of Korea\\
\{wonseok.choi, yhyuna, jongmin.kim, kevin5188, jeff1273, gajh\}@snu.ac.kr
}
}

\maketitle

\thispagestyle{empty}
\pagestyle{plain}
\setcounter{page}{1}

\begin{abstract}
Fully homomorphic encryption (FHE) enables secure computation on encrypted data, mitigating privacy concerns in cloud and edge environments.
However, due to its high compute and memory demands, extensive acceleration research has been pursued across diverse hardware platforms, especially GPUs.
In this paper, we perform a microarchitectural analysis of CKKS, a popular FHE scheme, on modern GPUs.
Focusing on the memory hierarchy, we demonstrate that dominant kernels remain bound by the on-chip L2 cache despite its high bandwidth, exposing a persistent inner memory wall beyond the conventional off-chip DRAM bottleneck.
Further, we reveal that the overall CKKS throughput is constrained by low per-kernel hardware utilization, caused by insufficient intra-kernel parallelism.
Motivated by these findings, we introduce \NAME, a set of complementary, memory-aware optimizations that improve cache efficiency and reduce runtime overheads.
\NAME achieves 1.45--1.83$\times$ performance improvements over a highly optimized baseline, Cheddar, across representative CKKS workloads.
On an RTX 5090, we reduce the bootstrapping latency for 32,768 complex numbers from 22.1ms to 15.2ms, and further to 12.8ms with additional algorithmic optimizations, establishing a new state-of-the-art GPU performance to the best of our knowledge.

\end{abstract}


\input{introduction}

\input{background}
\input{kernel}

\input{optimization}
\input{evaluation}

\input{discussion}
\input{related}
\input{conclusion}
\section*{Acknowledgments}
This work was supported by the Institute of Information \& Communications Technology Planning \& Evaluation (IITP) grant funded by the Korean government (MSIT) (RS-2025-02217656, RS-2021-II211343, IITP-2025-RS-2023-00256081).
Jongmin Kim and Hyesung Ji are with the Interdisciplinary Program in Artificial Intelligence, Seoul National University.
Jung Ho Ahn is the corresponding author. 

\bibliographystyle{IEEEtranS}
\bibliography{refs}

\end{document}

%% file: introduction.tex
\section{Introduction}

Fully homomorphic encryption (FHE) has long been considered the ``holy grail'' of cryptography~\cite{cacm-2010-holy-grail}, enabling arbitrary computation on encrypted data without ever exposing the underlying plaintext.
Among various FHE schemes, CKKS~\cite{asia-2017-ckks} has emerged as a promising solution for privacy-preserving machine learning (ML) due to its native support for fixed-point arithmetic and efficient plaintext structure for vector operations.
However, despite its theoretical promise, the computational overhead of FHE remains a major barrier to its practical deployment, often resulting in slowdowns of 2--6 orders of magnitude compared to plaintext execution~\cite{asplos-2026-cheddar}.

To bridge this performance gap, recent research has turned to GPUs as the primary acceleration platform~\cite{asplos-2026-cheddar,tches-2021-100x,isca-2025-neo,hpca-2025-warpdrive}.
GPUs offer massive parallelism and high memory bandwidth, making them a compelling platform for the large-degree polynomial arithmetic inherent to FHE.
State-of-the-art implementations, such as Cheddar~\cite{asplos-2026-cheddar}, have demonstrated remarkable speedups over baseline implementations based on CPUs~\cite{wahc-2021-hexl, wahc-2020-lattigo, wahc-2022-openfhe} or even FPGAs~\cite{hpca-2023-poseidon, hpca-2023-fab, hpca-2025-effact}.

Prior GPU studies have sought acceleration opportunities from various directions.
First, numerous studies have focused on optimizing compute-intensive FHE kernels, particularly the number-theoretic transform (NTT)~\cite{iiswc-2020-ntt, hpca-2023-tensorfhe, hpca-2025-warpdrive, isca-2025-neo}.
This line of work includes TensorFHE~\cite{hpca-2023-tensorfhe}, WarpDrive~\cite{hpca-2025-warpdrive}, and Neo~\cite{isca-2025-neo}, which repurpose otherwise idle data pipelines, such as tensor cores and FP64 units, on NVIDIA GPUs to accelerate integer operations required in FHE, thereby achieving additional arithmetic throughput.
On top of these compute optimizations, several studies~\cite{tches-2021-100x, asplos-2026-cheddar, ispass-2025-fideslib, isca-2025-neo, hpca-2025-warpdrive} apply kernel fusion to reduce memory transfers and kernel launch overheads.

Despite these efforts, the research community still lacks a systematic analysis of the current bottlenecks in GPU-based FHE execution and hence has remaining optimization opportunities.
We address this gap through a comprehensive microarchitectural study spanning individual kernels to full FHE workloads.

\emph{Our thorough analysis unveils a dominant memory wall in GPU-based FHE execution, driven by limitations in GPU memory hierarchy.} 
While prior studies~\cite{tches-2021-100x, iiswc-2020-ntt, hpca-2023-tensorfhe, ispass-2025-fideslib}, have focused on DRAM, we show that highly optimized GPU FHE libraries, such as Cheddar~\cite{asplos-2026-cheddar}, require a deeper analysis of the full memory hierarchy.
We identify that, even for the most compute-intensive operations such as NTT, prior compute-centric optimizations have culminated in highly tuned kernels that are constrained by memory bandwidth, despite the increased L2 bandwidth of modern GPUs.
Also, we highlight the limits of memory-centric optimizations, such as kernel fusion, stemming from small shared memory (L1 cache) capacity.
Ultimately, both bandwidth and capacity limitations, rooted in current memory technology, emerge as fundamental barriers to FHE performance on GPUs.

Based on this analysis, we introduce \emph{\NAME},\footnote{Theodosian draws its name from the Theodosian walls. It is used to
emphasize that the modern memory hierarchy, or ``memory wall,'' is comprised
of multiple tiers of on-chip and off-chip memory, each of which must be
considered in terms of its latency, capacity, bandwidth, and energy efficiency.} a set of microarchitectural optimizations that improve effective memory throughput and overall hardware utilization.
Rather than relying solely on further kernel-level tuning, we focus on managing data movement and execution to better utilize the GPU memory hierarchy.
First, we propose an L2-aware multi-polynomial caching strategy that batches operation sequences while keeping their working sets resident in the L2 cache, enhancing hardware utilization without incurring additional DRAM traffic.
Second, we introduce complementary pipelining, which executes DRAM-bound and L2-bound kernels together to improve aggregate memory throughput.
We also adopt CUDA Graphs~\cite{programming-guide} to eliminate kernel launch overhead by exploiting the static structure of FHE workflows.

Overall, \NAME delivers consistent performance improvements of 1.45--1.83$\times$ over Cheddar across representative CKKS workloads.
Our analysis indicates that the remaining room for improvement is fundamentally limited by the volume of global memory (L2 cache and DRAM) transactions, leaving a practical headroom of at most 1.73$\times$ under current algorithmic and hardware constraints.
This observation underscores the need for future work that moves the memory wall itself, rather than merely approaching it, through memory-aware cryptographic algorithms and hardware designs that deliver substantially higher on-chip memory bandwidth.

The key contributions of this work are as follows:
\begin{itemize}[leftmargin=*]
    \item We present a microarchitectural analysis of CKKS bootstrapping on modern GPUs, identifying that DRAM-only analysis is insufficient and hence the dominant bottleneck in each kernel lies in the global memory transactions rather than arithmetic throughput.
    \item We propose a set of memory-hierarchy-aware optimizations that improve effective memory throughput and overall hardware utilization, addressing the limited parallelism exposed by GPU kernels under practical CKKS parameters.
    \item \NAME achieves 1.45--1.83$\times$ performance improvements across representative workloads. In particular, for bootstrapping, recent algorithmic advances provide an additional 1.27$\times$ speedup.
\end{itemize}

%% file: background.tex
\section{Background}
\label{sec:back}

\subsection{The CKKS fully homomorphic encryption (FHE) scheme}
\label{sec:back:ckks}

When a server receives FHE ciphertexts, into which the plaintext data of a client are encrypted, it can evaluate functions over these ciphertexts, whereas conventional encryption schemes require decryption prior to computation.
An FHE scheme called CKKS~\cite{asia-2017-ckks} enables the client to encrypt a length-$\frac{N}{2}$ complex (or real) vector into a single ciphertext, allowing efficient $\frac{N}{2}$-way vectorized computations on encrypted fixed-point data.

Bootstrapping is essential for achieving unlimited computation in FHE.
FHE operations inherently accumulate noise (error), which eventually corrupts information in a ciphertext.
Bootstrapping evaluates the homomorphic decryption circuit to refresh the ciphertext, reducing the noise back to a manageable level~\cite{eurocrypt-2018-heaanboot}.
However, it is computationally expensive; bootstrapping accounts for a significant portion (e.g., 73\% for encrypted ResNet-20 inference on a GPU~\cite{asplos-2026-cheddar}) of the total execution time, making it the primary acceleration target.
Henceforth, we describe the computational characteristics of CKKS at a high level without delving into its cryptographic details~\cite{asia-2017-ckks, sac-2018-frns-ckks, eurocrypt-2018-heaanboot, rsa-2020-better}, which is beyond the scope of this work.

A CKKS ciphertext is expressed as a pair of polynomials $(\poly{a},\poly{b})$, and each polynomial consists of $L$ limbs~\cite{micro-2023-mad} $\poly{a}[i]_{0\le i < L}$.
A limb is also called a residue polynomial, having a degree of $N\!-\!1$.
Thus, each polynomial is represented as an $L\times N$ matrix of integer coefficients $\poly{a}[i][j]_{0\le i < L, 0\le j <N}$, each of which is typically 32-bit in contemporary implementations~\cite{asplos-2026-cheddar, wahc-2023-32bit}:
{
\setlength{\arraycolsep}{1.2pt}
\begin{equation*}
\poly{a}
=
\begin{bmatrix}
\poly{a}[0] \\
\vdots \\
\poly{a}[L\!-\!1]
\end{bmatrix}
=
\begin{bmatrix}
\poly{a}[0][0] & \cdots & \poly{a}[0][N\!-\!1] \\
\vdots & \ddots & \vdots \\
\poly{a}[L\!-\!1][0] & \cdots & \poly{a}[L\!-\!1][N\!-\!1]
\end{bmatrix}.
\end{equation*}
}

Each limb (row) $\poly{a}[i]$ is subject to a modulus $Q_i$, requiring reductions modulo $Q_i$ for every polynomial operation.
While modular additions are cheap, modular multiplications (mults) demand a complex sequence of integer operations, such as Montgomery reduction~\cite{1985-montgomery}.
$N\!=\!2^{16}$ is a typical value, fixed across polynomials, making each limb 0.25MB for 32-bit elements.
In contrast, $L$ varies across polynomials as polynomial operations such as BConv (\S\ref{sec:back:ops}) change $L$.
For a single polynomial with $N=2^{16}$, $L$ can take various values below 64, which is determined by security constraints~\cite{2021-standard}.

\subsection{Polynomial operations}
\label{sec:back:ops}

All high-level CKKS operations can be broken down into the following three types of polynomial operations~\cite{isca-2022-bts}.

\subsubsection{Number-theoretic transform (NTT)}

To facilitate a polynomial mult between \textpoly{a} and \textpoly{b}, which, without NTT, would require computing convolutions between the coefficients of each limb ($\poly{a}[i] * \poly{b}[i]$), NTT transforms each limb so that the convolutions become element-wise vector mults ($\odot$) due to the Fourier transform principles: $\poly{a}[i] * \poly{b}[i]=\mathtt{NTT}(\poly{a}[i])\odot\mathtt{NTT}(\poly{b}[i])$.
Using the fast Fourier transform algorithms~\cite{cooley-tukey, gentleman-sande}, NTT and inverse NTT (INTT) are individually decomposed into structured sequences of $\frac{N}{2}{\log_2 N}$ butterfly ($\mathtt{BFly}$) operations:
\begin{itemize}
\item $\mathtt{BFly}_\mathtt{NTT}(x,y,w) \rightarrow (x + w\cdot y, x - w\cdot y) \bmod Q_i$
\item $\mathtt{BFly}_\mathtt{INTT}(x,y,w) \rightarrow (x + y, w\cdot(x-y)) \bmod Q_i$
\end{itemize}
As NTT is performed for each limb, NTT exhibits \textbf{row-wise parallelism} and has a complexity of $\mathcal{O}(L N\log N)$.

\subsubsection{Base conversion (BConv)}

BConv is primarily a matrix-matrix mult (GEMM)~\cite{micro-2022-ark}.
For a polynomial \textpoly{a} of shape $L_\mathrm{in}\times N$,
BConv multiplies a constant table $\mathbf{T}$ of size $L_\mathrm{out}\times L_\mathrm{in}$ to produce an output polynomial with $L_\mathrm{out}$ limbs:
{
\setlength{\arraycolsep}{1.2pt}
\[
\mathtt{Red}
\left(
\left[
\begin{array}{c}
\cellcolor{blue!20}\mathbf{T}[0]\\
\vdots\\
\mathbf{T}[L_\mathrm{out}\!-\!1]\\
\end{array}
\right]
\cdot
\left[
\begin{array}{ccc}
\cellcolor{blue!20}\poly{a}[0][0] & \cdots & \poly{a}[0][N\!-\!1] \\
\cellcolor{blue!20}\vdots & \ddots & \vdots \\
\cellcolor{blue!20}\poly{a}[L_\mathrm{in}\!-\!1][0] & \cdots & \poly{a}[L_\mathrm{in}\!-\!1][N\!-\!1]
\end{array}
\right]
\right)
\]
}

\noindent When applying conventional tiled GEMM methods~\cite{cutlass-blog} to BConv, it is most efficient to partition \textpoly{a} by columns.
Partitioning $\mathbf{T}$ is less critical due to small $L$ values.
Thus, BConv exhibits \textbf{column-wise parallelism}.
Each multiplication requires modular reduction with respect to modulus $P_{i\in[0,L_\mathrm{out})}$, which differs from input modulus $Q_{j\in[0,L_\mathrm{in})}$.
As an optimization, modular reductions ($\mathtt{Red}$ in the above equation) are deferred until the end of GEMM~\cite{asplos-2026-cheddar}. 
For example, the accumulated values in the blue-highlighted region of the above equation are stored in a larger numeric space (e.g., 64-bit) to avoid overflow.
Then, $\mathtt{Red}$ operation is applied only once at the end.
Thus, BConv requires $L_\mathrm{in}\cdot L_\mathrm{out}\cdot N$ plain multiply-adds (MADs) and at least $L_\mathrm{out} \cdot N$ modular reductions.

\subsubsection{The rest}
All remaining polynomial operations exhibit \textbf{element-wise parallelism} and can be fully decomposed into ($L \times N$)-parallel integer operations.
For example, adding two polynomials (\textpoly{a} and \textpoly{b}) consists of $L \times N$ computations of $\poly{a}[i][j]+\poly{b}[i][j] \bmod Q_i$.
A special case is a data movement operation referred to as automorphism, which shuffles the order of the columns in a polynomial.
As its column index calculation is relatively cheap, automorphism can also be regarded as element-wise parallel.

\subsection{NVIDIA GPU execution model}
\label{sec:back:gpu}

Modern GPUs execute workloads by launching a massively parallel computational kernel, consisting of a grid of thread blocks.
Each block, typically consisting of 128--512 threads, executes on a single streaming multiprocessor (SM), which includes multiple data pipelines, such as FMA (fused multiply-add), ALU, and FP64.
For FHE, the most heavily utilized pipelines are FMA, used for integer mult, and ALU, used for other integer operations such as addition and comparison.
During kernel execution, each thread uses its own registers, and threads within a block can cooperate by storing intermediate results in software-controlled L1 cache space referred to as shared memory.
Shared memory is limited in size (e.g., 100 KB on the RTX 5090) and cannot be shared across SMs (and thus across thread blocks).

While the shared memory enables programmers to maximize data reuse within L1 cache for a single kernel, we cannot rely on L1 cache to retain data across multiple kernels.
Blocks cannot synchronize with each other within a kernel, and kernels execute independently.
Consequently, global memory, comprising L2 cache and DRAM, serves as the primary communication and storage boundary between kernels.
Any data that must be shared across blocks or persist across kernel invocations must be written back to global memory, which introduces significant bandwidth overhead.

%% file: kernel.tex
\section{Microarchitectural Analysis of Prior GPU FHE Acceleration}
\label{sec:boot}

\subsection{Why we need yet another analysis}
\label{sec:boot:motivation}

\begin{figure}[t]
    \centering
    \subfloat[L2 cache capacity]{\includegraphics[width=0.409\columnwidth]{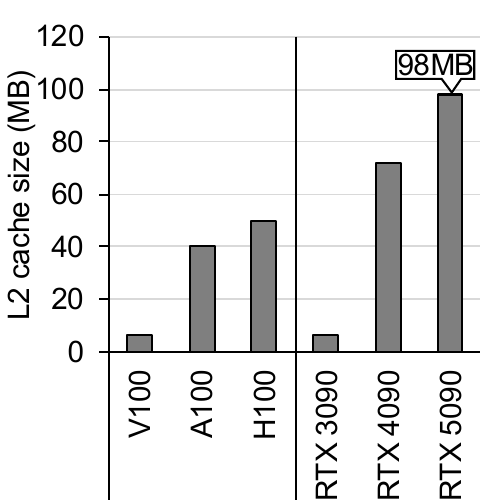}}\label{fig:gpu-spec:cachesize}
    \hfill
    \subfloat[Memory and compute throughput]{\includegraphics[width=0.56\columnwidth]{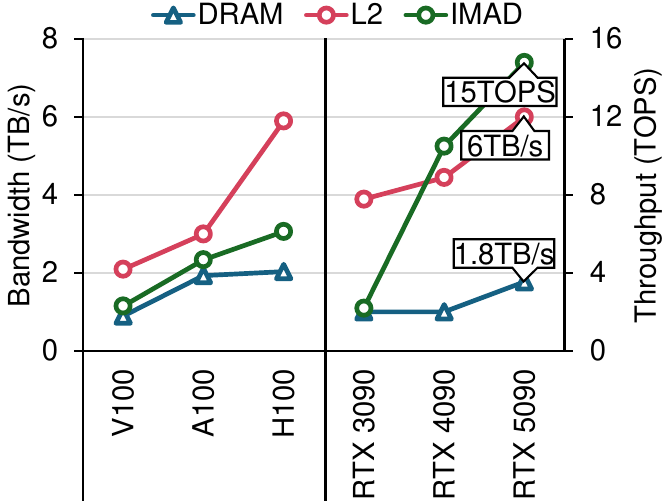}}\label{fig:gpu-spec:throughput}
    
    \caption{Comparing (a) L2 cache capacity, (b) empirical memory bandwidth, and empirical 32-bit integer multiply-and-add (IMAD) throughput across various NVIDIA GPUs. L2 bandwidth is measured by repeatedly reading and writing a data block smaller than the L2 cache size to saturate its bandwidth.
    }
    \label{fig:gpu-spec}
\end{figure}

While prior studies have identified DRAM bandwidth as the primary bottleneck for FHE on GPUs, recent advances in GPU hardware and FHE implementations demand a closer examination of the entire GPU memory hierarchy.
Since Jung et al.~\cite{tches-2021-100x} demonstrated that all major FHE kernels fall into the memory-bound region based on a DRAM-based roofline model on the V100 GPU, following studies~\cite{asplos-2026-cheddar, hpca-2025-warpdrive, hpca-2025-anaheim, isca-2025-neo, ispass-2025-fideslib} have focused on mitigating DRAM bandwidth pressure.
This DRAM-centric view becomes insufficient for modern GPUs due to significant shifts in the hardware landscape (see Fig.~\ref{fig:gpu-spec}).
RTX 5090 is the highest-end consumer GPU currently available, offering relatively high computational throughput and memory bandwidth among the tested GPU systems.
It features a 98MB L2 cache, more than 16$\times$ the capacity of the V100's 6MB.

Although the larger L2 cache enables more extensive data caching for FHE and allows us to leverage prior techniques, a substantial gap still remains between modern GPUs' memory hierarchy and existing FHE optimization methods:
\begin{itemize}[leftmargin=*]
    \item \textbf{Hardware proposals}~\cite{micro-2022-ark, isca-2023-sharp} typically assume massive scratchpad capacities of 180--512MB to eliminate the tight resource constraints of commodity hardware, limiting their applicability to commercial GPUs.
    \item \textbf{MAD}~\cite{micro-2023-mad}, the work most relevant to ours, reduces DRAM pressure via architecture-agnostic caching strategies. 
    However, MAD targets moderate cache sizes (6--32MB) and overlooks on-chip bandwidth constraints, limiting its applicability on modern GPUs with much larger but bandwidth-constrained L2 caches.
\end{itemize}

\begin{figure*}[t]
    \centering
    \subfloat[Kernel-level utilization analysis]{\includegraphics[width=1.25\columnwidth]{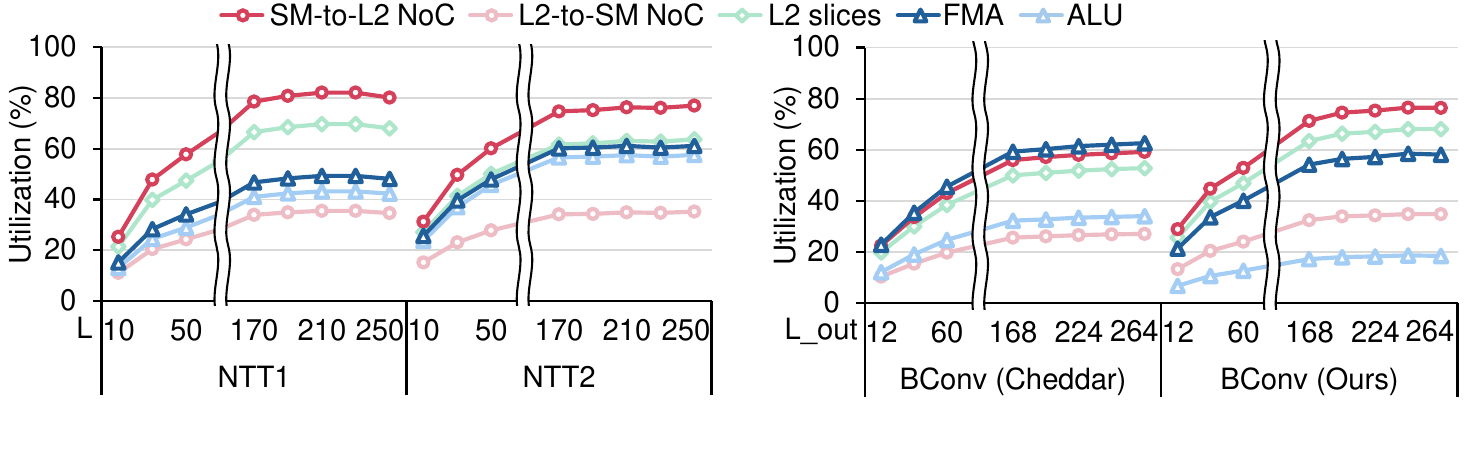}\label{fig:util:kernel}}
    \hfill
    \subfloat[Core Operation]{\includegraphics[width=0.2795\columnwidth]{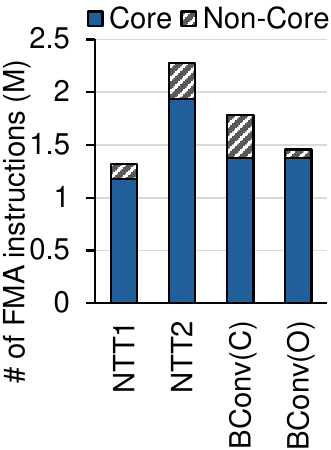}\label{fig:util:core}}
    \hfill
    \subfloat[Execution time]{\includegraphics[width=0.393\columnwidth]{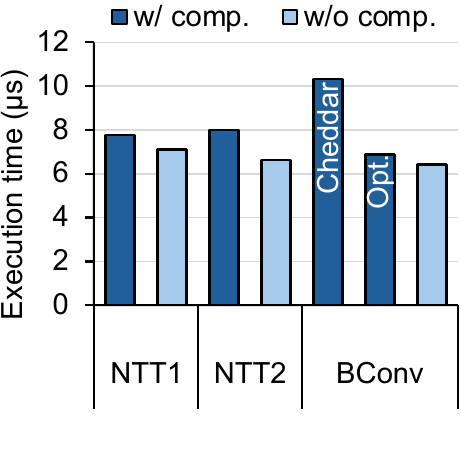}\label{fig:util:exec}}
    \caption{(a) Utilization rates of ALU, FMA, SM-to-L2 NoC, L2-to-SM NoC, and L2 slices for NTT kernels with $L$ varying from 10 to 250, and for BConv kernels with $L_{\mathrm{in}}=12$ and $L_{\mathrm{out}}$ varying from 12 to 264.
    BConv(C) and BConv(O) denote the original BConv in Cheddar~\cite{asplos-2026-cheddar} and our optimized BConv, respectively.
    We look beyond the typical $L$ range for a single polynomial ($L<64$), which will be made clear in \S\ref{sec:opt:parallel}.
    (b) FMA instruction breakdown into core (defined in \S\ref{sec:boot:motivation}) and non-core operations for NTT1, NTT2, BConv(C), and BConv(O).
    (c) Kernel execution time changes when the core operations are removed for $L = 48$ (NTT) and $(L_\mathrm{out}, L_\mathrm{in})=(48, 12)$ (BConv).
    An RTX 5090 GPU is used for the analysis.} 
    \label{fig:util}
    \vspace{-0.07in}
\end{figure*}

\begin{figure}[t]
    \centering
    \includegraphics[width=0.99\columnwidth]{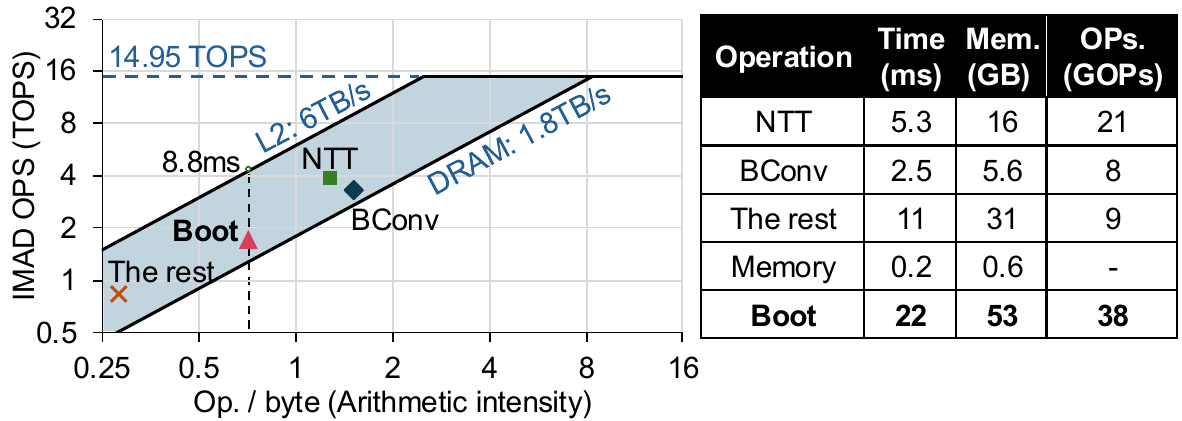}
    \caption{(Left) Roofline analysis of bootstrapping (Boot) and polynomial operations based on Cheddar~\cite{asplos-2026-cheddar} on RTX 5090. (Right) Time, global memory access, and core IMAD operation breakdown for bootstrapping. In addition to the operations, Boot includes kernel launch overhead.}
    \label{fig:boot}
\end{figure}

Our analysis bridges this gap by accurately assessing the capabilities of modern GPUs for FHE acceleration.
We used Cheddar~\cite{asplos-2026-cheddar}, the state-of-the-art GPU library, for the analysis.
The current L2 cache operates in a unique regime: it is large enough to cache and reuse several ciphertexts, yet still insufficient to contain complex FHE operation sequences (e.g., key-switching) entirely.
Further, the L2 bandwidth is only 2.1--4.0$\times$ higher than DRAM bandwidth, introducing a secondary bottleneck often overlooked in prior studies.

To quantify their implications, we first conducted a roofline analysis of bootstrapping, the most critical component of FHE, by counting the required global memory accesses, which pass through the L2 cache but may not reach DRAM, and the ``core'' 32-bit integer multiply-and-add (IMAD) operations.
The core operations include only computations on useful polynomial data, excluding auxiliary work such as index calculations.
Fig.~\ref{fig:util:core} shows that these core operations account for most FMA instructions, the primary data pipeline used by compute-intensive FHE kernels.
Based on the results shown in Fig.~\ref{fig:boot}, we make the following remarks:
\begin{itemize}[leftmargin=*]
    \item \emph{A DRAM-only analysis is insufficient}: the performance point lies above the DRAM roofline, indicating a need for a deeper look at the GPU memory hierarchy.
    \item Bootstrapping is bottlenecked by global memory bandwidth even when accounting for the higher L2 bandwidth. In fact, every kernel composing bootstrapping becomes L2-bandwidth-bound with kernel-level optimizations in \S\ref{sec:boot:kernel}.
    \item The L2 bandwidth on the RTX 5090 imposes a bootstrapping latency limit of 8.8ms, whereas Cheddar achieves 22.1ms, which is 2.5$\times$ above this bound. In \S\ref{sec:opt}, we present microarchitecture-aware optimizations designed to more closely approach this performance limit.
\end{itemize}

\subsection{Microarchitectural Characterization of FHE on GPUs}
\label{sec:boot:gpu}

To better understand the bottleneck indicated by Fig.~\ref{fig:boot}, we conduct a fine-grained characterization of polynomial operation kernels using hardware performance counters collected with NVIDIA Nsight Compute~\cite{nsight-compute}.
Specifically, we analyze compute pipeline utilization and L2 cache behavior to characterize how polynomial operation kernels stress the compute and memory subsystems.

\noindent \textbf{Compute characterization:} 
We focus on the FMA and ALU pipelines to quantify compute pipeline utilization (\S\ref{sec:back:gpu}).

\noindent \textbf{L2 cache characterization:}
NVIDIA GPUs partition the L2 cache into many independent slices, each serving as a proxy for DRAM.
Every global memory access from an SM must traverse the corresponding address-mapped L2 slice.
These L2 slices are connected to the SMs through multiple network-on-chip (NoC) paths, with SM-to-L2 and L2-to-SM transfers handled via partially decoupled networks.
As a result, L2 bandwidth utilization is jointly determined by the capacity of the SM-to-L2 and L2-to-SM NoCs and the read/write throughput of the L2 slices themselves~\cite{micro-2024-gpunoc,programming-guide}.
We follow the memory model defined in the NVIDIA profiling guide~\cite{profiling-guide}, which models the L2 cache as separate tag and data stages.
L2 slice utilization is measured as the maximum of \texttt{lts\_\_d\_sectors} and \texttt{lts\_\_t\_sectors}, because the larger value determines the L2 slice bottleneck.
Other L2 metrics, including those related to atomic operations, remain negligible in our analysis.

\subsection{Memory wall in polynomial operation kernels}
\label{sec:boot:kernel}

We first conducted a kernel-level analysis focusing on the computationally demanding NTT and BConv kernels based on Cheddar~\cite{asplos-2026-cheddar}.
For each kernel, we evaluated with multiple limb counts $L$, including values substantially higher (up to $264$) than those used by a single polynomial ($<64$).
These high-$L$ configurations correspond to cases with multiple parallel polynomial operations, which can arise even within single-ciphertext bootstrapping, as discussed in \S\ref{sec:opt:parallel}.
Varying $L$ over a wide range allows us to study how increasing levels of parallelism impact compute and memory utilization.

\subsubsection{NTT}

The NTT kernels in Cheddar are primarily limited by the constrained L2 write bandwidth, as indicated by their high SM-to-L2 NoC utilization.
NTT is split into two kernels, NTT1 and NTT2, to satisfy the constraints that will be discussed in \S\ref{sec:boot:fusion}.
Fig.~\ref{fig:util:kernel} shows that, for both kernels, the SM-to-L2 NoC is the hardware component most utilized, whose utilization saturates at around 80\% for large $L$ values.

NTT2 shows relatively higher compute utilization due to additional computations required for an optimization in Cheddar referred to as on-the-fly twiddle factor generation (OT~\cite{iiswc-2020-ntt}).
Without OT, NTT2 requires a larger number of twiddle factor loads ($w$ in $\mathtt{BFly}$ operations) compared to NTT1.
NTT1 requires only $N_1=\mathcal{O}(\sqrt{N})$ twiddle factors per limb, whereas NTT2 requires $N - N_1=\mathcal{O}(N)$ twiddle factors.
OT allows computing most of the twiddle factors for NTT with additional computations to alleviate the global memory access burden for NTT2.
Despite this optimization, NTT2 remains L2-bandwidth-bound as SM-to-L2 bandwidth continues to be the dominant bottleneck, even though its saturated FMA utilization increases by 21.3\% relative to NTT1.

\subsubsection{BConv}
Fig.~\ref{fig:util:kernel} shows that an optimized BConv implementation also becomes L2-bandwidth-bound. 
To understand our BConv optimization, we first examine Cheddar's modular reduction approach.
As described in \S\ref{sec:back:ops}, modular reductions can be deferred until the end of the GEMM operation in BConv.
However, because Cheddar uses moduli close to 32 bits ($Q_j \approx 2^{31}$), the accumulated value can easily exceed the capacity of a 64-bit register under typical settings.
For example, when $L_{\mathrm{in}} = 12$, the accumulated result can become as large as $3\cdot2^{64}$, necessitating costly intermediate reductions.
Thus, the original BConv kernel from Cheddar is highly compute-bound, heavily constrained by high FMA utilization.

We eliminate this computational overhead by refining the constant table $\mathbf{T}$.
For a given set of input moduli $Q_{j \in [0, L_\mathrm{in})}$ and output moduli $P_{i \in [0, L_\mathrm{out})}$, the elements of $\mathbf{T}$ are computed as follows:
$$\textstyle
\mathbf{T}[i][j]=(Q^*/Q_j)\bmod P_{i}\text{, where } Q^*=\prod_{j=0}^{L_\mathrm{in}-1}Q_j
$$
We can easily find a set of moduli satisfying the condition $\sum^{L_\mathrm{in}-1}_{j=0}\mathbf{T}[i][j]\cdot Q_j < 2^{64}$ for all $i$.
This condition is sufficient to guarantee overflow-free computation, thereby entirely removing the need for intermediate modular reductions.

This optimization substantially reduces the computational load of BConv.
As shown in Fig.~\ref{fig:util:core}, it cuts non-core operations by 18.5\%, which accelerates the overall BConv execution by 1.50$\times$ (Fig.~\ref{fig:util:exec}).
As a result, the performance bottleneck shifts from computation to memory bandwidth; Fig.~\ref{fig:util:kernel} demonstrates that our optimized BConv achieves high SM-to-L2 NoC utilization, indicating it has become strictly L2-write-bandwidth-bound.
Furthermore, this approach allows us to retain the same word-sized primes ($Q_j \approx 2^{31}$) as Cheddar, and the algorithmic refinement itself introduces no compromise to the original security level.

\subsubsection{The rest}

The remaining element-wise parallel kernels typically require very few computations per data element~\cite{asplos-2026-cheddar, hpca-2025-anaheim}, becoming bottlenecked by global memory bandwidth as shown by the roofline plot in Fig.~\ref{fig:boot}.

\noindent \textbf{Summary:}
Our analysis reveals a fundamental global memory bandwidth bottleneck present in every kernel of CKKS bootstrapping.
We also measured the execution time of each kernel with all the core operations (defined in \S\ref{sec:boot:motivation}) removed.
Removing the core operations yields only marginal performance improvements of 7\%--20\%, further confirming that these kernels are dominated by memory bandwidth rather than computation (see Fig.~\ref{fig:util:exec}).

Given this memory-bound behavior, a natural question is whether the memory traffic itself can be reduced.
All kernels are bottlenecked by the NoC bandwidth for SM-to-L2 writes, stemming from the lower L2 write bandwidth relative to reads~\cite{arxiv-2025-dissect-blackwell}.
However, each kernel writes only its final outputs (e.g., $L\times N$ for NTT and $L_\mathrm{out}\times N$ for BConv) to global memory, leaving no room to reduce the write volume.
Consequently, \emph{further kernel-level optimizations convey very little performance implication for FHE on modern GPUs.}

Fig.~\ref{fig:util:kernel} also shows that low-$L$ configurations suffer from severe hardware underutilization due to limited parallelism.
We address this issue in \S\ref{sec:opt:parallel}.

\subsection{L1 cache \& constraints of kernel fusion}
\label{sec:boot:fusion}

Kernel fusion is a widely adopted technique for reducing the global memory access by merging kernels that share input and/or output operands.
Prior work~\cite{hpca-2025-warpdrive,tches-2021-100x,asplos-2026-cheddar, isca-2025-neo, ispass-2025-fideslib} has introduced various kernel fusion methods tailored to FHE.

While Cheddar already includes the kernel fusion methods from prior work, further fusion opportunities are limited due to the following constraints:
\begin{enumerate}[leftmargin=*]
    \item For large degree $N$ such as $2^{16}$, NTT requires two kernels.
    \item It is impractical to fuse NTT, which uses row-level parallelism, with BConv, which uses column-level parallelism.
    \item Aside from NTT and BConv, element-wise parallel kernels can be fused with adjacent kernels or with one another. However, certain cases still fail to fuse effectively.
\end{enumerate}

The first limit arises from the small shared memory (L1 cache) capacity available per thread block.
While a single limb is sized 256KB, no existing GPU architectures support such a large shared memory space.
For example, RTX 5090 allows up to 100KB of shared memory per block.
As a result, NTT for a limb must be split across multiple blocks and the all-to-all data dependency in NTT necessitates inter-block data transfers.
This communication is handled by launching an additional NTT kernel to shuffle data and perform the remaining butterfly operations.
NVIDIA's cuFFT, a highly optimized GPU FFT library, also adopts a similar strategy~\cite{cufft}.

The second constraint also stems from the limited shared memory capacity.
When kernels exhibit different forms of parallelism, fusing them requires storing their combined working set in shared memory, which scales multiplicatively with their parallel dimensions.
For instance, fusing a row-parallel kernel (parallelized over $N$) with a column-parallel kernel (parallelized over $L$) would require accommodating an $L\times N$ matrix in shared memory to meet all the data dependencies, making such fusion impractical on current GPU architectures.

In contrast, element-wise kernels offer more flexibility.
Consider the following two cases:
\begin{itemize}
    \item $\poly{c} = NTT(\poly{a}_1) + \poly{a}_2 + \cdots \poly{a}_k$
    \item $\poly{c} = NTT(\poly{a}_1) + \cdots + NTT(\poly{a}_k)$
\end{itemize}
\noindent In the first case, fusion is feasible because the intermediate results can remain in registers or shared memory without increasing the required working set.
This approach has been successfully used by Jung et al.~\cite{tches-2021-100x} and Cheddar~\cite{asplos-2026-cheddar} to reduce global memory traffic.
In the second case, however, each NTT instance maintains its own working set in shared memory for later accumulation, causing the shared memory footprint to scale with $k$.
Such growth quickly exceeds the on-chip memory limits, making fusion impractical.

%% file: optimization.tex
\begin{figure*}[tb!]
    \centering
    \includegraphics[width=0.975\linewidth]{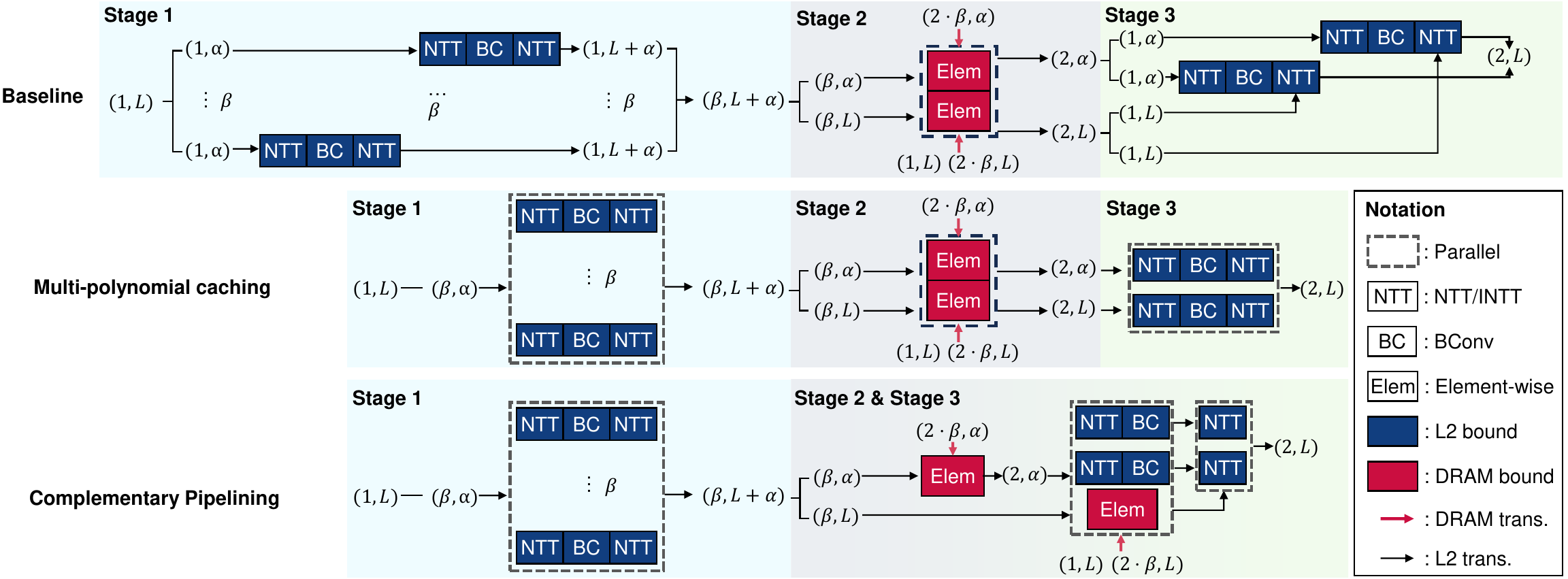}
    \caption{How the operational sequence of key-switching changes when multi-polynomial caching and complementary pipelining are applied. $(C, D)$ denotes $C$ polynomials each with $D$ limbs. Typical values for the parameters in the figure are $L=48$, $\alpha=12$, and $\beta=L/\alpha=4$.}
    \label{fig:keysw}
\end{figure*}

\begin{figure}[tb!]
    \centering
    \includegraphics[width=0.96\columnwidth]{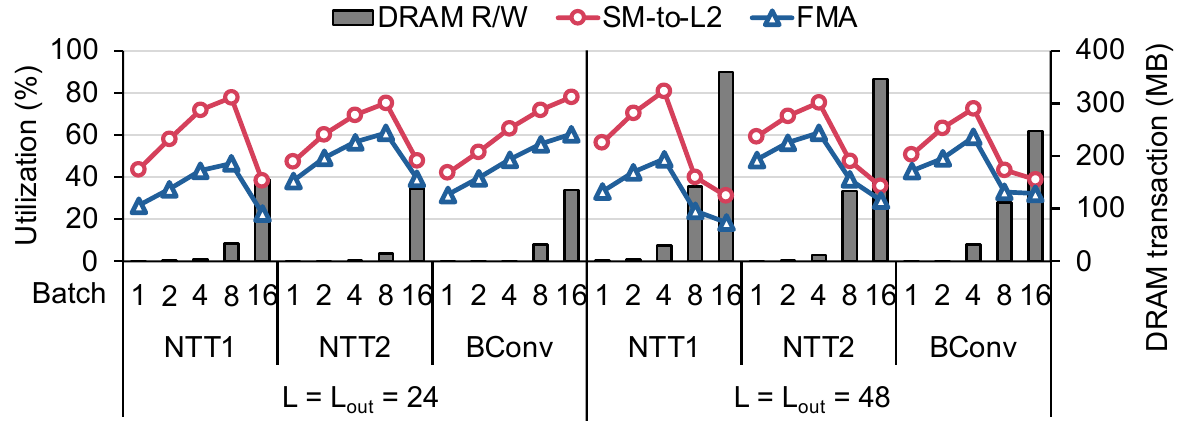}
    \caption{SM-to-L2 NoC utilization, FMA utilization, and the amount of DRAM transfers in
    NTT1, NTT2, and BConv kernels for batch sizes ranging from 1 to 16. Tested parameters are $L=24, 48$ for NTT and $(L_{\mathrm{out}},L_{\mathrm{in}})=(24,12),(48,12)$ for BConv. An RTX 5090 is used for the analysis.}
    \label{fig:batch}
\end{figure}

\section{Optimizations Under the Memory Wall}
\label{sec:opt}
The fundamental performance bound due to the limited global memory bandwidth and the kernel fusion constraints, discussed in \S\ref{sec:boot}, ultimately leads to a \emph{memory wall}.
Nevertheless, we may still boost the overall utilization of the memory system by carefully orchestrating CKKS operational sequences.
In this section, we present three microarchitecture-aware optimizations to improve GPU hardware utilization.
We use key-switching, a frequent operation sequence that dominates FHE runtime~\cite{tches-2021-100x,hpca-2023-fab,ispass-2024-ciflow}, as a key example to illustrate our optimizations.
Its workflow is shown in Fig.~\ref{fig:keysw}.

\subsection{Boosting hardware utilization with L2-aware batching}
\label{sec:opt:parallel}

Revisiting the kernel-level analysis in Fig.~\ref{fig:util}, the kernels’ L2 bandwidth utilization does not reach saturation for small $L$ values;
in fact, the saturation occurs for $L$ values far larger than those typically used for a single polynomial ($L<64$).

Batching can be used as a simple way to improve hardware utilization, inducing a similar effect as increasing $L$.
As shown in Fig.~\ref{fig:batch}, increasing the batch size raises the L2 utilization of each kernel to roughly 70--80\%.
Beyond a certain batch size, however, utilization drops due to a sharp increase in DRAM accesses.
Each batched operation enlarges the total working set and DRAM accesses become unavoidable once the combined footprint exceeds the L2 cache capacity.

While batching multiple ciphertexts, as in TensorFHE~\cite{hpca-2023-tensorfhe}, could expose additional parallelism, we focus on operational sequences found in single-ciphertext bootstrapping.
This avoids introducing an additional constraint on FHE usage.

MAD~\cite{micro-2023-mad} proposes caching strategies to reduce the DRAM pressure, albeit for relatively small cache capacities (6--32 MB).
Its $\mathcal{O}(\alpha)$-limb caching method, used in MAD's largest cache configuration, is also reflected in many GPU libraries, including Cheddar and FIDESlib~\cite{ispass-2025-fideslib}, for operational sequence organization.
$\mathcal{O}(\alpha)$-limb caching requires on-chip memory space larger than $2\alpha$ limbs, taking 6MB for typical parameters ($\alpha=12$ and 0.25MB per limb).
The baseline in Fig.~\ref{fig:keysw} shows how it handles key-switching;
multiple NTT-BConv-NTT sequences, each of which receives $\alpha$ input limbs, in key-switching are executed one by one, isolating the required data for each in an effort to keep the working set small enough to fit in on-chip memory.

However, with the larger cache capacities of modern GPUs, this strategy becomes suboptimal as the entire working set for each stage (up to 60MB for $\beta=4$ polynomials each with $L+\alpha=60$ limbs in stage 1) can now fit fully within the L2 cache, allowing inter-stage data reuse.
Therefore, instead of handling $\alpha$ limbs at a time to suppress the working set size growth, we can batch multiple NTT-BConv-NTT sequences to enhance the hardware utilization without increasing the DRAM accesses.

We generalize our approach with an L2-aware \textbf{multi-polynomial caching} method, which incorporates an additional batching dimension $B$ to increase GPU utilization while avoiding DRAM accesses.
This applies to any operation sequence as long as batching $B$ sequences keeps the total working set within the L2 cache capacity.
For key-switching, multi-polynomial caching is applied to stages 1 and 3 (see Fig.~\ref{fig:keysw}) with $B=\beta$ and $B=2$, respectively.

Furthermore, for smaller $L$ values, we can batch multiple key-switching sequences to make $B$ higher.
Even then, we determine $B$ based on the L2 cache capacity, distinguishing our approach from prior approaches such as TensorFHE~\cite{hpca-2023-tensorfhe}, which uses up to $B=128$ (enabled by multi-ciphertext bootstrapping) without considering the GPU memory hierarchy.

\subsection{Exploiting complementary memory bottlenecks}
\label{sec:opt:overlap}

Even with the large L2 cache capacity of modern GPUs, some data must still be fetched from DRAM due to poor reusability.
This includes key-switching hints~\cite{micro-2021-f1} shown as ($2\cdot\beta$, $\alpha$) and ($2\cdot\beta$, $L$) data in stage 2 of Fig.~\ref{fig:keysw}, where $(C, D)$ data denotes $C$ polynomials each with $D$ limbs ($\frac{CD}{4}$MB for typical parameters).

In such situations, \textbf{complementary pipelining} can be applied to improve effective bandwidth utilization.
The key idea is to pipeline a DRAM-bound kernel with an L2-bound kernel so that DRAM and the L2 cache are utilized more evenly, enhancing average memory bandwidth utilization.
How well the kernels are overlapped with pipelining depends on their aggregated L2 requirement; the benefits may be limited if the L2 cache components are already heavily utilized.

\begin{figure}
    \centering
    \includegraphics[width=0.9665\columnwidth]{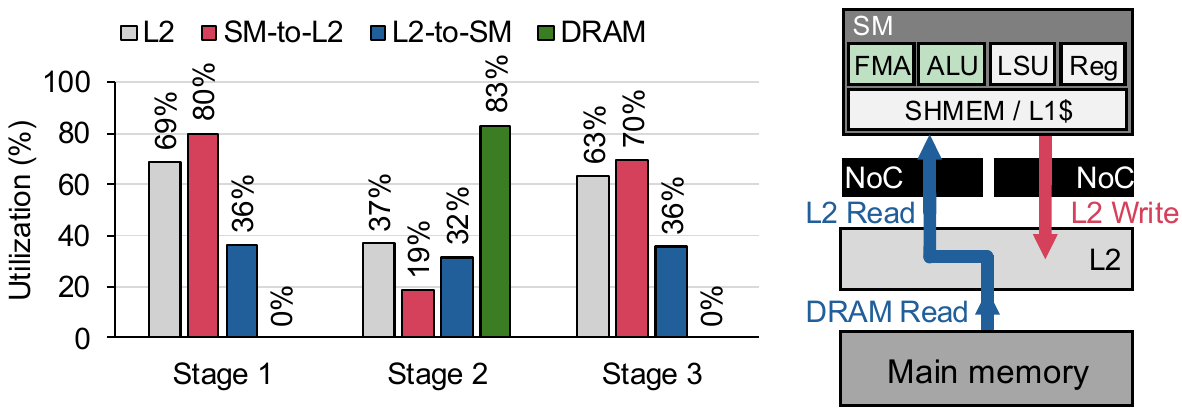}
    \caption{(Left) DRAM and L2 cache utilization for each stage of key-switching with an RTX 5090. (Right) Illustration of complementary pipelining on a simplified GPU architecture.}
    \label{fig:ks-stage-util}
\end{figure}

As an example, we illustrate how complementary pipelining is applied to key-switching.
Fig.~\ref{fig:ks-stage-util} shows DRAM and L2 cache utilization of each stage in key-switching.
Stages 1 and 3 consist primarily of NTT and BConv operations, both of which exhibit high SM-to-L2 NoC utilization and minimal DRAM traffic.
In contrast, stage 2 consists of an element-wise kernel, which additionally loads $(2\cdot\beta,\alpha)$, $(2\cdot\beta,L)$, and $(1, L)$ data, which can be as large as 135MB in total.
Due to this additional data, stage 2 exhibits high DRAM utilization but low L2 utilization, suggesting a DRAM bandwidth bottleneck.

We discover that stages 2 and 3 can be pipelined despite data dependencies.
Of the two stage-2 outputs, $(2,\alpha)$ and $(2,L)$ data, we only need the former at the start of stage 3.
Therefore, we split the original stage-2 kernel to produce the $(2, \alpha)$ data first and merge the $(2, L)$ computation with the subsequent stage-3 steps.
This allows DRAM transfers to be spread across stages 2 and 3, giving them more time to complete.

The decoupled NoC structure (\S\ref{sec:boot:gpu}) allows these two stages to run concurrently with modest L2 cache interference.
By co-executing these kernels with complementary memory bandwidth requirements, we can increase both DRAM and L2 bandwidth utilization, improving throughput beyond what is achievable by optimizing each kernel in isolation.

\subsection{Adopting CUDA Graphs}
\label{sec:opt:cuda-graph}
In addition to limited parallelism within individual kernels, hardware utilization is further reduced by the large number of fine-grained kernel launches in bootstrapping.
In Cheddar, a single bootstrapping operation launches 1,543 kernels, where fusing these kernels is constrained as described in \S\ref{sec:boot:fusion}.
Each launch incurs a non-negligible latency overhead of roughly 2--5µs due to context switching and runtime scheduling~\cite{isca-2021-BlockMaestro,sosp-2025-KTransformers}.
Given that each kernel’s actual execution time itself is on the order of microseconds (see Fig.~\ref{fig:util:exec}), this launch overhead directly reduces effective hardware utilization by increasing idle periods between kernels. 

To address this issue, we leverage CUDA Graphs~\cite{programming-guide}, motivated by the observation that FHE workloads are entirely static. 
Because computation on encrypted data cannot involve data-dependent branches, FHE execution graphs are fixed, making them a natural fit for a graph-based execution model.
With kernels as nodes and data dependencies as edges, the entire computation graph (a CUDA Graph) is submitted to the GPU in a single CPU launch, greatly reducing the kernel launch overhead.

%% file: evaluation.tex
\section{Evaluation}
\label{sec:eval}

We apply all our memory-hierarchy-aware optimizations in \S\ref{sec:boot} and \S\ref{sec:opt} to Cheddar~\cite{asplos-2026-cheddar}, resulting in a highly optimized GPU implementation that we call \textbf{\NAME}.
We evaluate \NAME by comparing it with state-of-the-art GPU studies~\cite{asplos-2026-cheddar, discc-2024-fhe-cnn, heaan-latest, hpca-2025-warpdrive, ispass-2025-fideslib}.

\subsection{Experiment setup}
\label{sec:eval:setup}

We used parameter sets with 128-bit security throughout all evaluations.
Except for \S\ref{sec:vs_heaan2}, all experiments adopt the parameter set in Table~\ref{tab:parameter}.
All experiments were conducted on an RTX 5090 unless otherwise specified.
We evaluate the following widely-used~\cite{hpca-2023-tensorfhe, hpca-2025-warpdrive, asplos-2026-cheddar, isca-2025-neo} FHE workloads that require bootstrapping:
\begin{itemize}[leftmargin=*]
    \item \textbf{Boot}~\cite{acns-2022-sparseboot}: Bootstrapping of a complex length-$2^{15}$ vector. After a single bootstrapping operation, the ciphertext supports evaluating a multiplicative circuit of depth 13 ($Lv_{\mathrm{eff}}=13$) for the default parameter set in Table~\ref{tab:parameter}, leaving 19 limbs ($L=19$). 
    \item \textbf{HELR}~\cite{aaai-2019-helr}: Logistic regression model training with 32 iterations, evaluated on a batch of 1024 MNIST~\cite{2012-mnist} grayscale images ($14\times14$). We report per-iteration execution time.
    \item \textbf{ResNet20}~\cite{icml-2022-resnet}: Inference of the ResNet20 model, a 20-layer convolutional neural network (CNN) with ReLU activations, on a single CIFAR-10 image ($32\times32\times3$).
\end{itemize}

\begin{table}[t]
    \centering
    \caption{Default evaluation parameters. $\mathtt{dnum}$ denotes decomposition number~\cite{rsa-2020-better}. $H_d$ ($H_s$) is the Hamming weight of dense (sparse) secret~\cite{rsa-2020-better}. $PQ$ is the maximum modulus. $\Delta$ determines precision, where high $\Delta$ results in higher computational cost.}
    \label{tab:parameter}
    \begin{tabular}{c|ccccccc}
    \toprule
       Param & $N$ & $L$ & $\mathtt{dnum}$ & $\log_2{PQ}$ & $H_d$ & $H_s$ & $\Delta$  \\
    \midrule
        Value & $2^{16}$ & $\le 48$ & 4  & $\le 1746$ & $2^{15}$ & $2^5$ & $2^{40}$ \\ 
    \bottomrule
    \end{tabular}
\end{table}

\begin{table}[t]
\centering
\caption{Workload execution times using \NAME compared to HEaaN-GPU~\cite{heaan-latest, discc-2024-fhe-cnn}, Cheddar~\cite{asplos-2026-cheddar}, WarpDrive~\cite{hpca-2025-warpdrive}, and FIDESlib~\cite{ispass-2025-fideslib}}
\label{tab:workload}
\begin{tabular}{l|c|ccccc}
\toprule
\multirow{2}{*}{GPU} & \multirow{2}{*}{Implementation} & Boot & HELR & ResNet20 \\
& & (ms) & (ms/it) & (ms) \\
 \midrule
\multirow{5}{*}{A100 80GB} & HEaaN-GPU & 171 & - & 8580\\ 
& WarpDrive & 121 & 113 & 5880 \\
& FIDESlib & 337 & - & - \\
& Cheddar & 40.0 & 51.9 & 1320\\
 & \textbf{\NAME}  & 31.8 & 37.5  & 966\\
 \midrule
\multirow{3}{*}{H100 80GB} & FIDESlib & 294 & - & -\\ 
& Cheddar & 31.2 &  40.7 & 1050\\
& \textbf{\NAME} & 23.6 & 26.5 & 740 \\
\midrule
\multirow{3}{*}{RTX 5090} & FIDESlib & 147 & - & - \\
 & Cheddar  & 22.1 & 25.9 & 720\\
 & \textbf{\NAME}  & 15.2  & 14.1 & 467 \\
 \midrule
 H100 SXM\textsuperscript{\dag} & \textbf{\NAME} & 17.8 & 21.3 & 577 \\
 \midrule
 H200 SXM & \textbf{\NAME} & 16.1 & 19.8 & 528 \\ 
 \midrule
B200 SXM & \textbf{\NAME} & 14.0 & 18.0 & 472 \\
\bottomrule
\end{tabular}
\begin{itemize}
    \item[\dag] H100 SXM is the high-performance server variant with higher compute throughput and memory bandwidth than the H100 80GB (PCIe) version.
\end{itemize}
\end{table}

\subsection{FHE Workload}
\label{sec:eval:workload}

\NAME achieves 1.45--1.83$\times$ overall speedups over Cheddar on the RTX 5090 (see Table~\ref{tab:workload}).
Specifically, bootstrapping latency is reduced to 15.2ms, which is 1.45$\times$ faster than Cheddar under the same parameter set.
For two higher-level workloads, ResNet20 and HELR, which both incorporate bootstrapping yet exhibit distinct computation patterns, we achieve 1.54$\times$ (ResNet20) and 1.83$\times$ (HELR) speedups.
A detailed analysis of their workload-level behaviors is presented in \S\ref{sec:eval:opt:workload}.

We additionally evaluate \NAME on multiple GPU platforms, achieving consistent improvements of 1.26--1.54$\times$ over Cheddar on the A100 80GB and the H100 80GB.
Moreover, compared to FIDESlib~\cite{ispass-2025-fideslib}, state-of-the-art among open-source GPU FHE libraries, we achieve 9.7--12.5$\times$ lower bootstrapping latency for the same $Lv_{\mathrm{eff}}$.
As a reference for high-end GPU configurations, we also include performance results for the H100, H200, and B200 SXM platforms.

\subsection{Ablation study of our optimizations}
\label{sec:eval:ablation}

\begin{table}
    \centering
    \caption{Maximum memory footprint and global memory transactions across different key-switching stages (Stage 1, 2, 3) for various parameter sets $(L,\alpha,\beta)$}
    \label{tab:ks-param}
    \begin{tabular}{ccccccc}
    \toprule
        \multirow{2}{*}{$(L,\alpha,\beta)$} & \multicolumn{2}{c}{Max. mem. footprint} & \multicolumn{3}{c}{Glob. mem. trans.} \\ 
   & Stage 1 & Stage 3 & Stage 1 & Stage 2 & Stage 3  \\
    \midrule
        $(48,12,4)$ & 62.9MB & 50.3MB & 352MB & 233MB & 201MB\\
        $(24,12,2)$ & 18.9MB & 25.2MB & 101MB & 81.8MB & 107MB\\
        $(12,12,1)$ & 6.30MB & 12.6MB & 31.5MB & 34.6MB & 69.2MB\\
    \bottomrule
    \end{tabular}
\end{table}
\begin{figure}
    \centering
    \subfloat[Ablation study (key-switching)\label{fig:ks-ablation}]{\includegraphics[width=0.55\columnwidth]{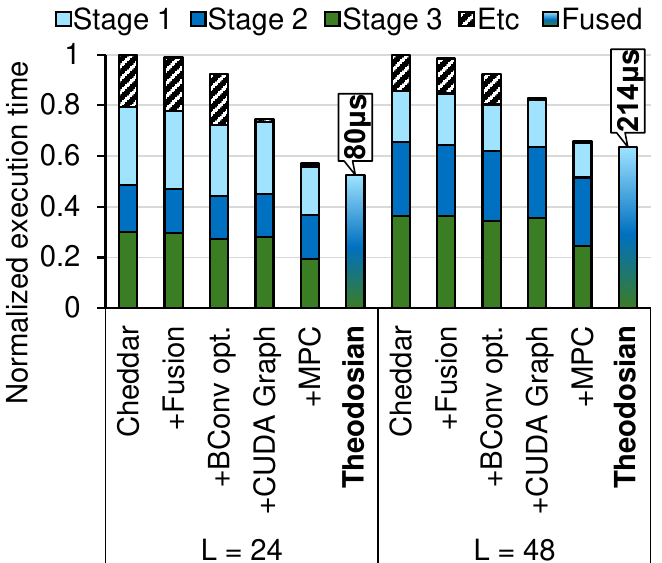}}
    \subfloat[Batching effect\label{fig:ks-batching}]{\includegraphics[width=0.4105\columnwidth]{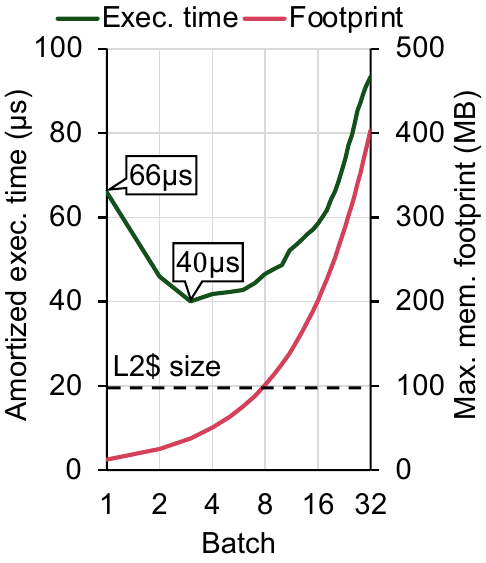}}
    \caption{(a) Execution time of key-switching when incrementally applying additional kernel fusion (+Fusion), BConv kernel optimization (+BConv opt.), CUDA Graph (+CUDA Graph), multi-polynomial caching (+MPC), and complementary pipelining (\NAME) to Cheddar for $L=24$ and $L=48$. (b) Amortized execution time and maximum memory footprint when processing multiple key-switching sequences in a batch for $L=12$. }
    \label{fig:ablation}
\end{figure}
\begin{figure}[t]
    \centering
    \subfloat[Ablation study (Boot)\label{fig:workload-boot}]{\includegraphics[width=0.3966\columnwidth]{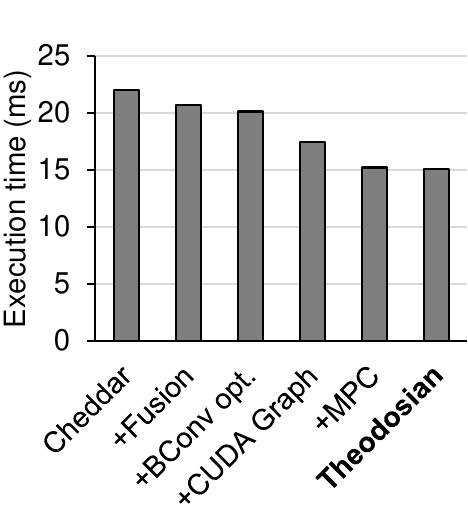}}
    \hspace{0.03\columnwidth}
    \subfloat[Workload breakdown\label{fig:workload-breakdown}]{\includegraphics[width=0.486\columnwidth]{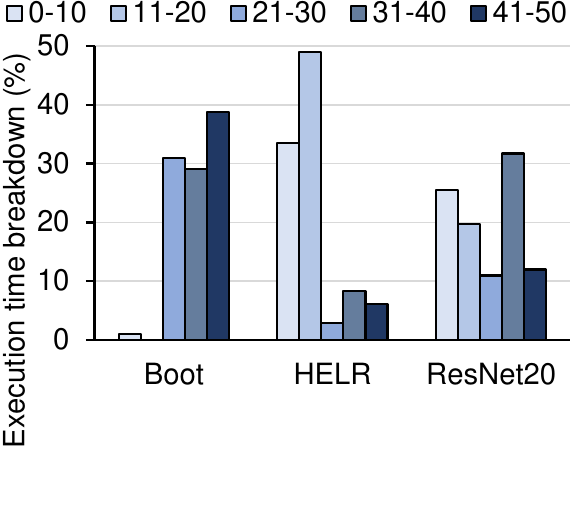}}
    \caption{(a) Ablation study of bootstrapping while incrementally applying our optimizations. (b) Execution time breakdown by limb count ($L$) for bootstrapping, HELR, and ResNet20, illustrating the fraction of total execution time across different $L$ ranges.}
    \label{fig:workload}
\end{figure}
We evaluate the effectiveness of our individual optimization techniques on key-switching and the workloads. 
We quantify the incremental contribution of each technique through an ablation study that progressively adds additional kernel fusion, BConv kernel optimization (\S\ref{sec:boot:kernel}), and our three microarchitecture-aware optimizations described in \S\ref{sec:opt}.
We used the three parameter configurations in Table~\ref{tab:ks-param} for key-switching experiments.

\subsubsection{Key-switching evaluation}

\NAME achieves 1.90$\times$ and 1.57$\times$ speedups over Cheddar for key-switching with $L=24$ and $L=48$, respectively (see Fig.~\ref{fig:ks-ablation}), where $L=48$ is the maximum number of limbs under our parameter set and $L=24$ is half that value.
The larger improvement at $L=24$ results from the lower hardware utilization at small $L$, which allows our optimizations to have a greater impact.

To understand where the gains come from, we break down the optimizations for $L=48$.
Although we explored additional kernel fusion beyond Cheddar, it resulted in little performance benefit because Cheddar already heavily fuses the key-switching sequence.
In contrast, our BConv kernel optimization (\S\ref{sec:boot:kernel}) provides an 8\% improvement over Cheddar. 
Using CUDA Graphs nearly eliminates miscellaneous overheads (e.g., kernel launch), which account for 9\% of the execution time. 
Building on this, our multi-polynomial caching optimization yields additional speedups of 1.46$\times$ for stage 1 and 1.36$\times$ for stage 3 of key-switching.
These results correspond to 70\% and 72\% of the ideal latency predicted from the required global memory traffic in Table~\ref{tab:ks-param} and the peak L2 bandwidth (6TB/s).
Finally, complementary pipelining further improves latency by 4\%.
For $L=24$, the miscellaneous portion rises to approximately 20\% due to finer-grained kernels, making CUDA Graph particularly effective and contributing to a larger overall speedup compared to $L=48$.

We also test the effect of batching for $L=12$, which makes a polynomial small enough to enable batching multiple key-switching sequences.
Table~\ref{tab:ks-param} lists the maximum memory footprints of stage 1 and stage 3, which are well below the L2 cache capacity for $L=12$.

To isolate the utilization benefit of batching, we intentionally avoid sharing common data (e.g., key-switching hints and twiddle factors) across batched sequences, even though such sharing would reduce memory traffic.
As a result, even as the batch size increases, the number of memory transactions per operation remains unchanged. 

Under this setting, batching improves the amortized key-switching execution time by up to 1.65$\times$ at $L=12$ (see Fig.~\ref{fig:ks-batching}).
However, once the batched footprint exceeds the L2 cache capacity of 98MB, performance degrades accordingly.

\subsubsection{Workload evaluation}
\label{sec:eval:opt:workload}
Bootstrapping shows a slightly different trend from key-switching: our sequence-aware kernel fusion improves performance by 7\% over Cheddar, while complementary pipelining yields only marginal gains, partly because Cheddar already applies hoisting~\cite{crypto-2018-linear,eurocrypt-2021-efficient} to reorder key-switching operations.
The remaining optimizations continue to provide meaningful performance improvements as shown in Fig.~\ref{fig:workload-boot}.

Bootstrapping primarily operates at large $L$, whereas ResNet20 and HELR exhibit different patterns.
Fig.~\ref{fig:workload-breakdown} presents the execution time breakdown according to the limb count ($L$) for each workload.
In particular, HELR spends 82\% of its execution time at small $L$ (0--20), where hardware utilization is lower.
This also reflects that HELR is a low-bootstrapping workload~\cite{asplos-2026-cheddar}, with bootstrapping accounting for a relatively small portion of its total execution time.
As a result, HELR is more sensitive to our microarchitectural optimizations and achieves more pronounced end-to-end performance gains.
In contrast, ResNet20 is largely dominated by bootstrapping~\cite{asplos-2026-cheddar} and performs relatively little work at small $L$ compared to HELR, leading to overall improvements similar to standalone bootstrapping.

%% file: discussion.tex
\section{Implications on Future Work}
\label{sec:vs_heaan2}

Compared with the 8.8ms bootstrapping latency bound imposed by 53GB of global memory accesses (Fig.~\ref{fig:boot}), the 15.2ms latency achieved by \NAME is only 1.73$\times$ higher.
This indicates that further gains from optimizations such as compute-focused software optimizations or hardware support for additional instructions are limited to at most 1.73$\times$ under this CKKS bootstrapping configuration on RTX 5090.
Moreover, this bound is derived under an aggressive assumption that even low-reuse data, such as plaintexts and key-switching hints, are served from L2 rather than DRAM (e.g., via ideal prefetching).
Given that we have already explored the practical limits in kernels and operational sequence organization through detailed microarchitectural analysis, further improvements would be challenging to achieve.

Instead of attempting to approach this memory wall, future work could focus on shifting the memory wall itself through cryptographic algorithm enhancements and better workload mapping to FHE settings.
For example, a recent proprietary CKKS library, HEAAN2~\cite{heaan2} reports a faster bootstrapping latency of 14.17ms on the RTX 5090.
While the exact algorithms used in HEAAN2 are not fully known, an optimization to reduce the memory transfers for polynomials is mentioned in its documentation.

We also attempted to incorporate the latest algorithmic optimizations, including OverModRaise~\cite{iacr-2025-overmodraise}, subring key encapsulation~\cite{asia-2025-subring}, and weight plaintext compression~\cite{ccs-2025-wpc}, into \NAME.\footnote{We adopt OverModRaise1, which saves one multiplicative level during bootstrapping. In addition, subring key encapsulation reduces the number of key-switching operations, while weight plaintext compression reduces plaintext loads. For further details on these techniques, we refer readers to the original papers.}
With these optimizations and appropriate parameter selection, we reduce global memory accesses from 53GB to 44GB, significantly alleviating memory bandwidth pressure.
The algorithmic enhancements result in an additional 1.27$\times$ speedup (+Algo. opt. in Table~\ref{tab:vs_heaan}) for \NAME, even outperforming HEAAN2 by 1.11$\times$.

However, even after the algorithmic optimizations, bootstrapping remains fundamentally constrained by data movement.
The total global memory transaction volume of 44GB imposes a latency limit of 7.3ms on the RTX 5090, leaving little headroom for further improvements.

\setlength{\tabcolsep}{3pt}
\begin{table}[t]
\centering
\caption{Bootstrapping comparison of Cheddar~\cite{asplos-2026-cheddar}, \NAME, and \NAME with additional algorithmic optimizations (+Algo. opt.), with HEAAN2~\cite{heaan2}.}
\label{tab:vs_heaan}
\begin{tabular}{lccccccc}
\toprule
 & $\log_2{PQ}$ & Boot prec.$\textsuperscript{\dag}$ & $\Delta$ & $H_d$ & $H_s$ & $Lv_{\mathrm{eff}}$ & Exec. time \\
\midrule
Cheddar & 1,711 & -18.57 & $2^{35}$ & 1024 & 32 & 15 &  23.18ms\\ 
\NAME & 1,711 & -18.57 & $2^{35}$ & 1024 & 32 & 15 &  16.21ms \\ 
+Algo. opt. & 1,711 & -18.70 & $2^{35}$ & 1024 & 32 & 15 & \textbf{12.75}ms \\
\midrule
HEAAN2 & 1,714 & -18.65 & $2^{38}$ & 1024 & 32 & 15 & 14.17ms \\
\bottomrule
\end{tabular}
\begin{itemize}
    \item[\dag]$\log_2$ of the maximum error averaged over 100 iterations
\end{itemize}
\end{table}
\setlength{\tabcolsep}{6pt}

Hardware-based solutions will require improvements in on-chip memory bandwidth; however, scaling bandwidth for a large on-chip structure, such as the L2 cache on modern GPUs, is inherently challenging.
Although prior ASIC FHE accelerator proposals~\cite{isca-2022-bts, isca-2022-craterlake, micro-2022-ark, isca-2023-sharp, isca-2025-fast, micro-2025-hawk, micro-2024-trinity, micro-2024-ufc, hpca-2025-effact, micro-2023-mad} achieve high bandwidth from massive on-chip memory, this is feasible largely because the memory is partitioned across compute units and does not support the fully random access required of a GPU L2 cache.

Finally, our optimizations are not specific to NVIDIA GPUs and are applicable to other GPU architectures as well.
AMD's MI300X, for example, scales out to a large number of compute units and pairs that with a huge 256MB last-level cache, while the local memory capacity corresponding to shared memory remains small at 64KB per compute unit~\cite{amd-mi300x-benchmark}.
In such a design space, our approach is aligned well as it focuses on improving cache efficiency and effective bandwidth utilization.

%% file: related.tex
\section{Related Work}
\label{sec:related}

Previous work on GPU-based FHE has explored multiple strategies to better leverage available hardware resources.
Jung et al.~\cite{tches-2021-100x} present one of the earliest GPU implementations of CKKS bootstrapping, reporting over 200$\times$ speedups over single-threaded CPU baselines by leveraging extensive kernel fusion optimizations.
TensorFHE~\cite{hpca-2023-tensorfhe} exploits NVIDIA tensor cores to accelerate NTT operations, while WarpDrive~\cite{hpca-2025-warpdrive} integrates both tensor cores and conventional CUDA cores for NTT.
NEO~\cite{isca-2025-neo} offloads HE kernels to underutilized FP64 pipelines, and it also studies an alternative key-switching method called KLSS~\cite{crypto-2023-klss}.
While KLSS can reduce computational complexity, it incurs 1.21--1.58$\times$ more global memory transactions than our baseline key-switching method~\cite{rsa-2020-better} across different limb sizes~\cite{isca-2025-neo}.

Cheddar~\cite{asplos-2026-cheddar}, which we have extensively used as our baseline, incorporates various kernel fusion techniques and exploits 32-bit integer arithmetic enabled by a GPU-friendly parameter selection methodology.
The use of small-integer arithmetic for FHE has also been studied in prior work~\cite{wahc-2023-32bit, isca-2023-sharp, asplos-2024-bitpacker, ccs-2025-grafting}.
FIDESlib~\cite{ispass-2025-fideslib} aims to build an open-source GPU FHE library on top of OpenFHE~\cite{wahc-2022-openfhe}.
A recent proprietary CKKS library, HEAAN2~\cite{heaan2}, reports a state-of-the-art bootstrapping latency of 14.17ms via algorithmic optimizations.
Although the algorithmic details are unavailable, we were able to surpass HEAAN2 (\S\ref{sec:vs_heaan2}) through microarchitecture-aware GPU optimizations and algorithmic enhancements.

Cerium~\cite{arxiv-2025-cerium} proposes a compiler-based framework with multi-GPU support for large-scale encrypted inference.
In contrast, this work focuses on single-GPU microarchitectural characterization and optimization, leaving multi-GPU scaling strategies beyond the scope of this paper.
Although Cerium targets similar workloads, it adopts a different algorithm; thus, we exclude it from our direct performance evaluation to ensure a fair comparison.

Meanwhile, GME~\cite{micro-2023-gme} introduces architectural modifications to an AMD GPU to better support polynomial operations.
Also, numerous ASIC-based FHE studies~\cite{isca-2022-bts, isca-2022-craterlake, micro-2022-ark, isca-2023-sharp, isca-2025-fast, micro-2025-hawk, micro-2024-trinity, micro-2024-ufc, hpca-2025-effact, micro-2023-mad,hpca-2026-crophe} have been proposed.
For example, FAST~\cite{isca-2025-fast} proposes a custom ASIC accelerator for FHE, achieving a remarkable bootstrapping latency of 1.38ms.

However, attaining such performance requires hundreds of MBs of on-chip memory that also provides over an order of magnitude higher bandwidth compared to the L2 cache of current GPUs.

HERACLES~\cite{isscc-2026-heracles} is a programmable FHE accelerator SoC fabricated in Intel 3 CMOS~\cite{vlsitc-2025-intel3}, featuring an 8192-way SIMD vector compute engine and a native polynomial ISA supporting a wide range of ring parameters, with SDK-level support for multiple FHE schemes.
HERACLES demonstrates the practical feasibility of large-scale hardware acceleration for FHE.

Several works~\cite{tches-2025-velofhe,tches-2025-xlk,cal-2024-tfhe,hpca-2026-peregrine} focus on the TFHE~\cite{jc-2020-tfhe} scheme, a fully homomorphic encryption scheme distinct from CKKS. 
From a performance analysis perspective, TFHE differs substantially from CKKS in both arithmetic structure and execution characteristics. 
In particular, TFHE employs much smaller polynomial degrees (e.g., $2^9$--$2^{11}$), leading to smaller working sets and different data-movement behavior. 
Its bootstrapping mechanism also differs fundamentally from that of CKKS, requiring distinct optimization strategies.

%% file: conclusion.tex
\section{Conclusion}
\label{sec:conclusion}

In this work, we have conducted a microarchitectural study of CKKS on modern GPUs.
We discovered that, even with today’s high-bandwidth L2 cache, the dominant primitive kernels remain fundamentally memory-bandwidth-bound, revealing a persistent memory wall in GPU-based FHE execution.
Motivated by this observation, we proposed three memory-hierarchy-aware optimizations to increase effective memory throughput, achieving 1.45--1.83$\times$ speedups across representative CKKS workloads.
Nevertheless, our analysis indicates that the remaining room for improvement is fundamentally constrained by the volume of global memory transactions, leaving at most a 1.73$\times$ headroom for further gains.
This observation points to future work that moves the memory wall itself through memory-aware cryptographic algorithms and hardware designs with higher on-chip memory bandwidth.